\documentclass[aps,onecolumn,preprint,a4paper]{revtex4}

\usepackage{graphicx}

\begin{document}

\newcommand{\aap}{Astron. \& Astroph.}
\newcommand{\mnras}{Month.Not.Roy.Astr.Soc. }
\newcommand{\apjs}{Astrophys.J.Suppl.Ser.}

\newcommand{\hh}{$\mathrm{H_2}$}
\newcommand{\hho}{$\mathrm{H_2O}$}
\newcommand{\hdo}{$\mathrm{HDO}$}
\newcommand{\ddo}{$\mathrm{D_2O}$}
\newcommand{\mhho}{\mathrm{H_2O}}
\newcommand{\spb}{\sigma^{\text{p.b.}}}
\newcommand{\ecol}{E_{\text{coll.}}}
\title{\emph{Ab initio} computation of the broadening of water rotational lines  by molecular hydrogen}
\author{Laurent Wiesenfeld}
\affiliation{Laboratoire d'Astrophysique de Grenoble, CNRS/Universit\'e Joseph-Fourier, Grenoble, France}
\author{Alexandre Faure}
\affiliation{Laboratoire d'Astrophysique de Grenoble, CNRS/Universit\'e Joseph-Fourier, Grenoble, France}

\begin{abstract}
Theoretical cross sections for the pressure broadening by hydrogen of
rotational transitions of water are compared to the latest available
measurements in the temperature range 65$-$220~K. A high accuracy interaction potential is employed in a full close coupling calculation. A good
agreement with experiment is observed above $\sim$80~K while the sharp drop observed
experimentally at lower temperatures is not predicted by our
calculations. Possible explanations for this discrepancy include the
failure of the impact approximation and the possible role of
ortho-to-para conversion of H$_2$.
\end{abstract}

\maketitle
Water is a prominent molecular component of interstellar matter. It
has been observed in most astrophysical environments, both in gas and
solid phase, as the most abundant polyatomic molecule
\cite{2005SSRv..119...29C}. Understanding water spectra is a key to
the general thermodynamical budget of astrophysical objects, because
of the many allowed spectral transitions, in mm, sub-mm and infra-red
regions of the spectrum. Also, the chemical and even exo-biological
importance of water cannot be understated. A detailed comprehension of
water physical chemistry in various environments is a major goal of
the newly launched Herschel Space Observatory (HSO).

In order to extract information from a molecular rotational spectrum,
it is essential to model its excitation scheme. Indeed, at low
densities, some lines may appear in absorption, some others in
emission and there is no \emph{a priori} reason for the molecule under
scrutiny to be in thermodynamical equilibrium with the main neutral
gas, \hh. Hence retrieving physical information from spectral lines is
only possible with a careful modeling of the interaction of the water
molecule with its photonic and molecular environments. Obtaining such
models has been a continuous effort in three directions: studies of
radiative transfer mechanisms, scattering theory, and very
importantly, laboratory experiments capable of testing various
theories and models pertaining to collisions \cite{eltzur:book}. Many
comparisons between experiments and theory are nowadays underway, in
order to put the water-hydrogen interaction on a firm ground
 \cite{2010a:yang,2010b:yang,belpassi10,2010A&A...517A..13D}.

In this communication, we wish to show the first results of a fully
\textsl{ab initio} computation of pressure broadening cross sections,
$\spb (T)$, based on a high precision potential energy surface (PES)
for the water-hydrogen interaction
\cite{2005JChPh.122v1102F,2008JChPh.129m4306V}(hereafter V08), as
compared to the very recent experiments of pressure broadening at low
temperatures by Dick et al.
\cite{2009JQSRT.110..619D,2010PhRvA..81b2706D}(hereafter DDP10). To
our knowledge, it is the first time that such a full \emph{ab initio}
pressure broadening quantum calculation is performed for a non-linear
molecule in interaction with another molecule. Other comparisons were
very successful for simpler symmetries
\cite{2008PCCP...10.5419T,2007JMoSp.246..118T,2005JChPh.122r4319R,1991JChPh..95.3888G,1991JChPh..94.1346G}.
%


Water being an asymmetric rotor, the rotational levels are usually
denoted as $j_{\kappa_a\kappa_c}$ or $(j, \tau)$ where $j$ is the
rotational quantum number associated with the angular momentum and
($\kappa_a, \kappa_c$) (projections of $j$ along inertia axis)  are pseudoquantum numbers, and $\tau = \kappa_a - \kappa_c$. The rotational constants of
\hho\ are taken at $A=27.88063134\, \mathrm{cm}^{-1}$,
$B=14.52176959\,\mathrm{cm}^{-1}$,
$C=9.277708381\,\mathrm{cm}^{-1}$. Rotational constant of \hh\ is
taken at $B=60.853\,\mathrm{cm}^{-1}$.
\par 

Following the experiments of DDP10, we compute here the pressure
broadening of the two spectral transitions connecting the ground
states of water: the para 1113~GHz line ($1_{11}\leftarrow 0_{00}$)
and the ortho 556~GHz line ($1_{10}\leftarrow 1_{01}$). All our
calculations are based on the V08 water-hydrogen full-dimensional PES
which was obtained by combining standard CCSD(T) calculations with
elaborate explicitly correlated CCSD(T)-R12 calculations. As in
Dubernet et al. \cite{2006A&A...460..323D}, we have employed the
rigid-body version of the V08 PES obtained by averaging the
full-dimensional PES over the ground vibrational states of the
monomers. Full details can be found in \cite{2008JChPh.129m4306V}.

The broadening of a rotational spectral line because of collision with
a buffer gas has been studied theoretically and experimentally for a
long time and theory is by now well established. The very general impact
approximation states that collision times are much shorter than time
between collisions. Within that approximation, which we discuss later,
it has been shown that the pressure broadening cross section for the
transition from initial state $i$ to final state $f$, at temperature
$T$, $\sigma^{\text{p.b.}}_{f \leftarrow i}(T)$, may be expressed by
closed expressions based on the transition matrix $\mathbf{T}$
\cite{1958PhRv..112..855B,1975AdChP..33..235B}.

For a broadening coefficient $\Gamma_{f\leftarrow i }(T)$, in
frequency/pressure units, Baranger \cite{1958PhRv..111..481B} defined
the pressure broadening cross section at energy $E$,
$\sigma^{\text{p.b.}}(E)$ as
\begin{equation}
\Gamma_{f\leftarrow i }(T) = 1/2 \,\left<nv\sigma^{\text{p.b.}}_{f\leftarrow i }(E)\right>_{ T\text{\ Boltzmann av.}}  
\end{equation}
where $i$ and $f$ are the initial and final states of the transition,
$n$ is the density of the observed molecule, and $v$ is the relative
velocity of water and hydrogen. $E$ is the collision kinetic
energy. Hence it is possible to define a Boltzmann averaged
$\sigma^{\text{p.b.}}_{f\leftarrow i }(T)$:
\begin{equation}
\sigma^{\text{p.b.}}_{f\leftarrow i }(T) = \frac{1}{T^2}\int \sigma^{\text{p.b.}}_{f\leftarrow i }(E)  \mathrm{e}^{-E/T}\,E\,\mathrm{d}E .
\label{eq:ave}
\end{equation}

Two equivalent ways have been proposed to calculate
$\sigma^{\text{p.b.}}_{f\leftarrow i }(E)$, and consequently the
averaged $\sigma^{\text{p.b.}}_{f\leftarrow i }(T)$. Following
Baranger \cite{1958PhRv..112..855B}, Schaefer \& Monchik
\cite{1992A&A...265..859S,1987JChPh..87..171S} and Green
\cite{1977CPL:green}, we have, for a rotational transition of
\hho\ $\left(j_f,\tau_f \leftarrow j_i,\tau_i\right)$, assuming that
\hh\ remains in an unchanged $j_2$ state:
\begin{widetext}
\begin{eqnarray}
\sigma^{\text{p.b.}}(j_f \tau_f j_2 \leftarrow j_i \tau_i j_2; E) & = & \frac{\pi}{k^2}\frac{1}{2j_2+1}\;\sum_{J_i J_f}\;\;\sum_{l, l',j_{12}j'_{12},\bar{j}_{12}\bar{j}'_{12}}
X(J_i,J_f,j_i,j_f,l, l',j_{12}j'_{12},\bar{j}_{12}\bar{j}'_{12}) \label{eq:sumav}\\ 
 & & \times\left<j_2l_0j_{12}j_i\tau_i\left|\mathbf{T}^{J_i}(E_i)\right|j_2l'_0j'_{12}j_i\tau_i\right>\left<j_2l_0\bar{j}_{12}j_f\tau_f\left|\mathbf{T}^{J_f}(E_f)\right|j_2l'_0\bar{j}'_{12}j_f\tau_f\right>^*\nonumber 
\end{eqnarray}
\end{widetext}
In Eq.(\ref{eq:sumav}), $j_{12}, j'_{12}$ are the angular quantum numbers
numbers resulting from the coupling of angular momenta $j_2$ and $j_{i}$ / $j_f$ ( for example, $\left|j_2-j_i\right| \leq j_{12},\,j'_{12}\leq (j_2+j_i)$). $l_0, l'_0$ are the orbital quantum
numbers. $\mathbf{T}^J(E)=1-\mathbf{S}^J(E)$ is the transition matrix,
at total angular momentum $J$. The $X(.)$ function groups all angular
coupling coefficients and parity sign terms; it is explicit in \cite{1987JChPh..87..171S}. $E_i$ and $E_f$ are the
two initial and final \emph{total} energies ($E_i\neq E_f$), corresponding to the same
kinetic energy, $(\hbar k)^2/2\mu$, $\mu$ being the collision reduced
mass and $k$ the momentum. 
A similar equation relevant for \hho -
\hh\ coupling may be found in Monchik \cite{1987JChPh..87..171S},
Eq.~(1). It should be noted the $1/(2j_2+1)$ factor in front of
Eq.(\ref{eq:sumav}), lacking in \cite{1977CPL:green} is necessary,
as underlined for example in \cite{2001PCCP....3.3924T}.

Baranger, followed by many authors, proceeded to compute the $\spb(E)$
in a different, yet equivalent way, with help of the optical
theorem. We have, with nearly the same notations as
\cite{1958PhRv..112..855B}:
\begin{eqnarray}
\sigma_{f\leftarrow i}^{\text{p.b.}}(E)& =  &\frac12\left[\sum_{f'}\sigma^{\text{inel.}}_{f'\leftarrow i}(E) + \sum_{i'}\sigma^{\text{inel.}}_{f\leftarrow i'}(E)
   \right] \nonumber \\ & & +
\int \left|f_i(\Omega; E)-f_f(\Omega; E)\right|^2 \mathrm{d}\Omega
\label{eq:optical}
\end{eqnarray}
In Eq.(\ref{eq:optical}), $\sigma^{\text{inel.}}(E)$ are ordinary
\emph{inelastic} cross-sections, $i'$ and $f'$ being all levels
connected to $f$ or $i$ at kinetic energy $E$. The $f_{i}(\Omega; E)$,
$f_f(\Omega; E)$ are the differential \emph{elastic} scattering
amplitudes, for the initial and final states, which must be set to
interfere before taking the modulus and integrating over all
scattering angles $\Omega$.

It must be strongly underlined that both approaches are
equivalent. Quite often, since inelastic cross sections or rate
coefficients are made available in the literature, in order to have an
estimate of $\sigma_{f\leftarrow i}^{\text{p.b.}}(E)$, equation
(\ref{eq:optical}) is truncated: only the first two terms, the
inelastic cross sections, are used, yielding sometimes to reliable
results \cite{1977JCPgreen:1409} and sometimes not
\cite{2005JChPh.122r4319R}, depending on the structure of the
scattering amplitudes (see below).

In order to calculate the $\mathbf{T}$ matrix elements of
Eq.~(\ref{eq:sumav}), we performed a full quantum close coupling
scattering computation with help of the Molscat program
\cite{molscat}. The $\mathbf{T}$ matrix elements were subsequently
summed at each kinetic energy to get the relevant $\sigma(E)$
cross-sections, inelastic and pressure-broadening. We separately
computed collisions of the four symmetry types: (para/ortho
\hho)--(para/ortho \hh). Parameters of the Molscat calculations are
similar to the ones used previously\cite{2010A&A...517A..13D,
2009A&A...497..911D,2006A&A...460..323D,2003A&A...408.1197G,2002A&A...390..793D},
with the following rotational basis sets: para-\hh : $j_2 =0,2$;
ortho-\hh : $j_2=1$. Ortho and para \hho: $j_1\leq 5,6,7$, for
increasing collision energy. For $E < 20\,\mathrm{cm}^{-1}$, the
hybrid modified log-derivative/Airy propagator of Manolopoulos and
Alexander was used; above that energy, the diabatic modified
log-derivative method of Manolopoulos was used \cite{molscat}. We
checked convergence with respect to basis set size, maximum range of
radial integration and size of step in the radial propagation.  The
collision energy range was $0.5 \leq E \leq 350 \,\mathrm{cm}^{-1}$,
with increments $\Delta E$ small enough to roughly describe
resonances. Decrease of the $\Delta E$ step did not change
significantly the averaged $\sigma^{\text{p.b.}}_{f\leftarrow i }(T)$,
which was obtained by a numerical integration of
Eq.~(\ref{eq:ave}). At collision energies $ E > 350\;\rm cm^{-1}$,
$\spb(E)$ remains essentially flat and was therefore extrapolated as a
constant.

\begin{figure}[htp]
\begin{center}
\includegraphics[angle=0,width=0.4\textwidth]{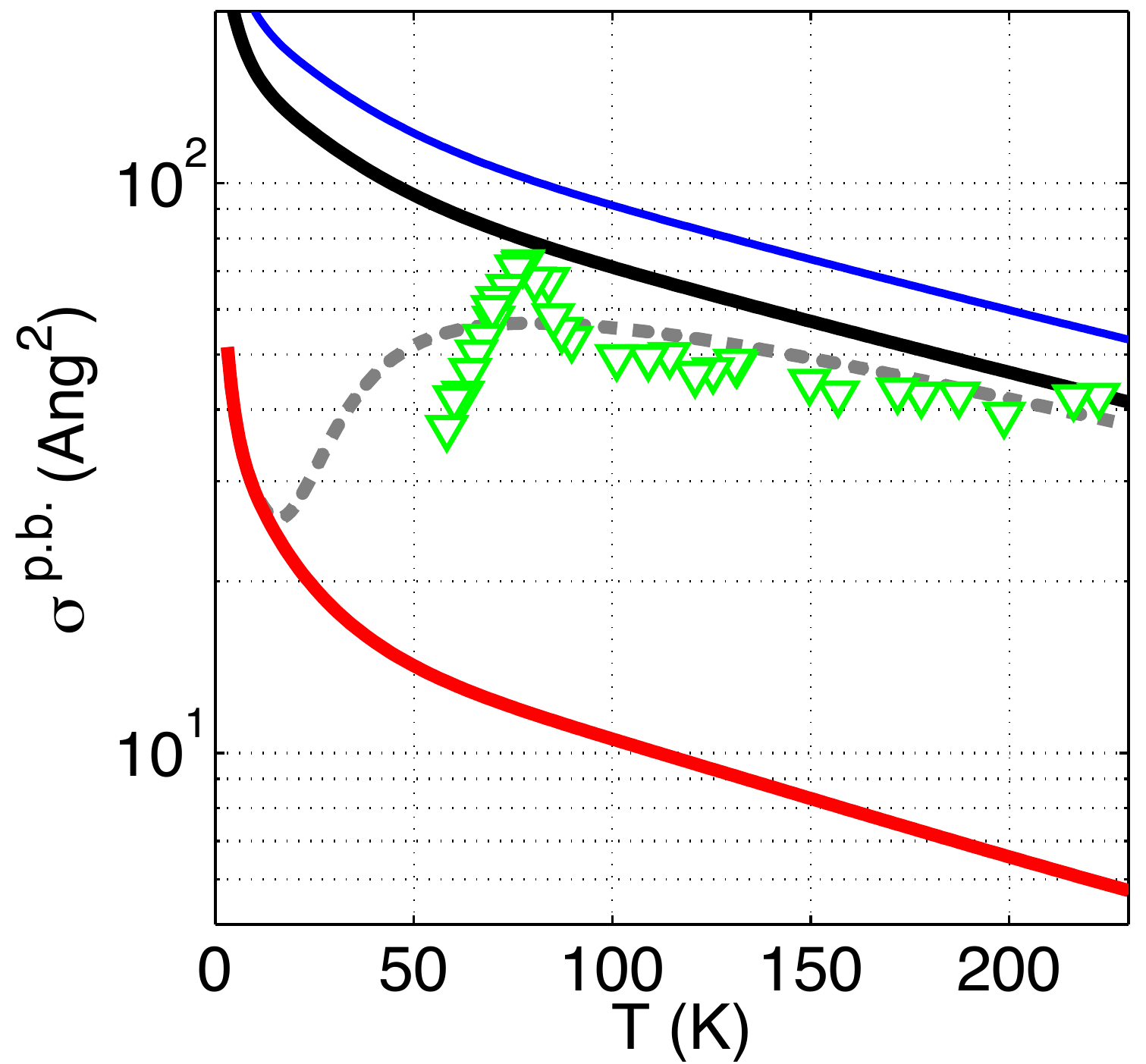}
\caption{Pressure broadening cross-sections for the transition at 556
  GHz, ortho-H$_2$O, $1_{10}\leftarrow 1_{01}$. Theory: Red line,
  below, pure para-H$_2$ collisions; blue line above, pure ortho-H$_2$
  collisions. Black continuous line, OPR value, 3; grey dashed line OPR
  value at equilibrium for each $T$. Green triangles, experimental
  values taken from \protect\cite{2010PhRvA..81b2706D}.}
\label{fig:985}
\end{center}
\end{figure}

\begin{figure}[htp]
\begin{center}
\includegraphics[angle=0,width=0.4\textwidth]{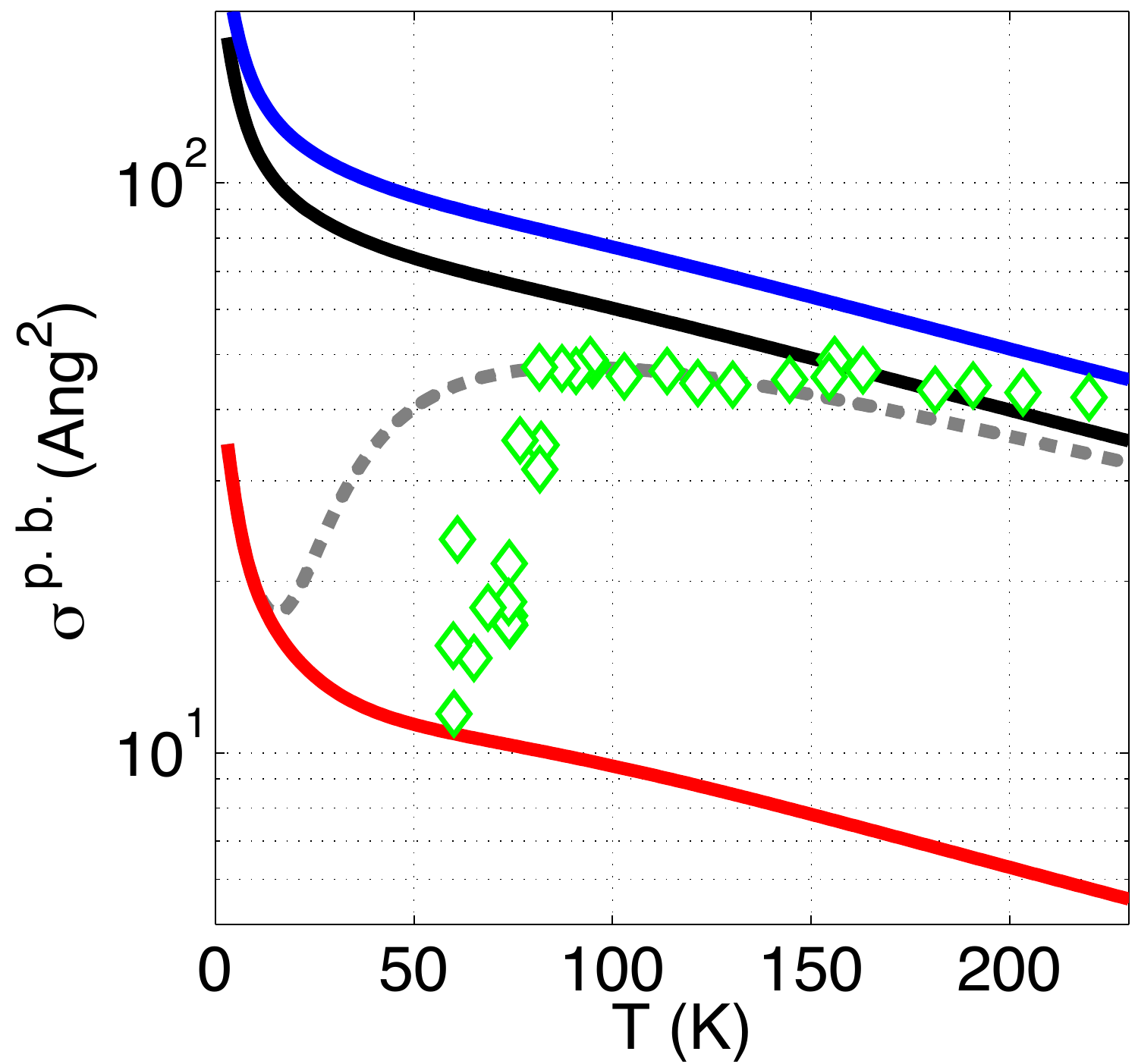}
\caption{Pressure broadening cross sections for the transition at 1113
  GHz, para-H$_2$O, $1_{11}\leftarrow 0_{00}$. Same caption as for figure
  \protect\ref{fig:985}.}
\label{fig:1100}
\end{center}
\end{figure}

Results are presented in figures \ref{fig:985} and \ref{fig:1100}. In
addition to the pure ortho H$_2$ (blue lines) and para H$_2$ (red
lines) cross sections, we added two possibilities for the
ortho-to-para ratio (OPR) of H$_2$: the solid black lines suppose a
``normal'' OPR value of 3, as expected in the DDP10 experiment. The
grey solid lines suppose that the thermodynamical equilibrium OPR
value is reached at each temperature $T$, as if ortho to para
transitions were possible. In both figures, the results of DDP10 are
shown as green open symbols, with their exact values provided courtesy
of B. Drouin.

Several points may be seen by inspection of figures~\ref{fig:985}
and~\ref{fig:1100}. If we assume an OPR value of 3, we see that the
theory vs. experiment agreement is very good (within 30\%) for $T
\gtrsim 80\,\rm K$. Let us recall that there are no adjustable
parameters involved in the simulation, except for the OPR value. This
agreement should come at no surprise, since recent analogous
calculations on branches of Raman spectra show similar successes
\cite{2008PCCP...10.5419T,2009JMoSp.256...17T,2007JChPh.126t4302G,2001PCCP....3.3924T}. Both
these calculations and ours make use of state-of-the-art \emph{ab
  initio} PES as well as fully converged close-coupling
calculations. Within the impact approximation valid at these
temperatures and moderate densities ($ n\lesssim 1 $~amagat), Baranger
formalism \cite{1958PhRv..112..855B}, equation~(\ref{eq:sumav}), is
essentially exact for nonoverlapping lines. The whole uncertainty that
arises is due either to inadequacies of the impact approximation,
which is to be ruled out here at $T>80$~K (see for example
\cite{1975AdChP..33..235B} or \cite{hartmann:book} for a discussion)
or else to imprecision of the PES. We see thus that we actually test
the PES by comparing experimental pressure broadening with careful
quantum calculations, in the relevant density and temperature ranges.

Pressure broadening cross sections, at these intermediate
temperatures, are sensitive to the overall shape of the PES, and
especially to the isotropic part. This may be understood in two
ways. First, the $\mathbf{T}$ matrix elements actually used in
Eq.(\ref{eq:sumav}) are elastic in the rotational quantum numbers,
thus non-zero also for an isotropic potential energy surface. Second,
this is confirmed by the good quality of pressure broadening
coefficients obtained by approximate semi-classical methods, where the
impinging trajectory of the perturber is totally decoupled from the
tensorial nature of the molecule/molecule electromagnetic interaction
\cite{hartmann:book,2008JQSRT.109.2857Y,2006MolPh.104.2701N}. In
retrospect, we understand that approximating the full
Eq.(\ref{eq:sumav}) by the purely inelastic terms of
Eq.(\ref{eq:optical}) may be correct in certain cases, but this should
at least be carefully checked. An analogous point has been made
earlier, see \cite{2001PCCP....3.3924T}.
 
\begin{figure}[tb]
\begin{center}
\includegraphics[width=0.4\textwidth,angle=-90]{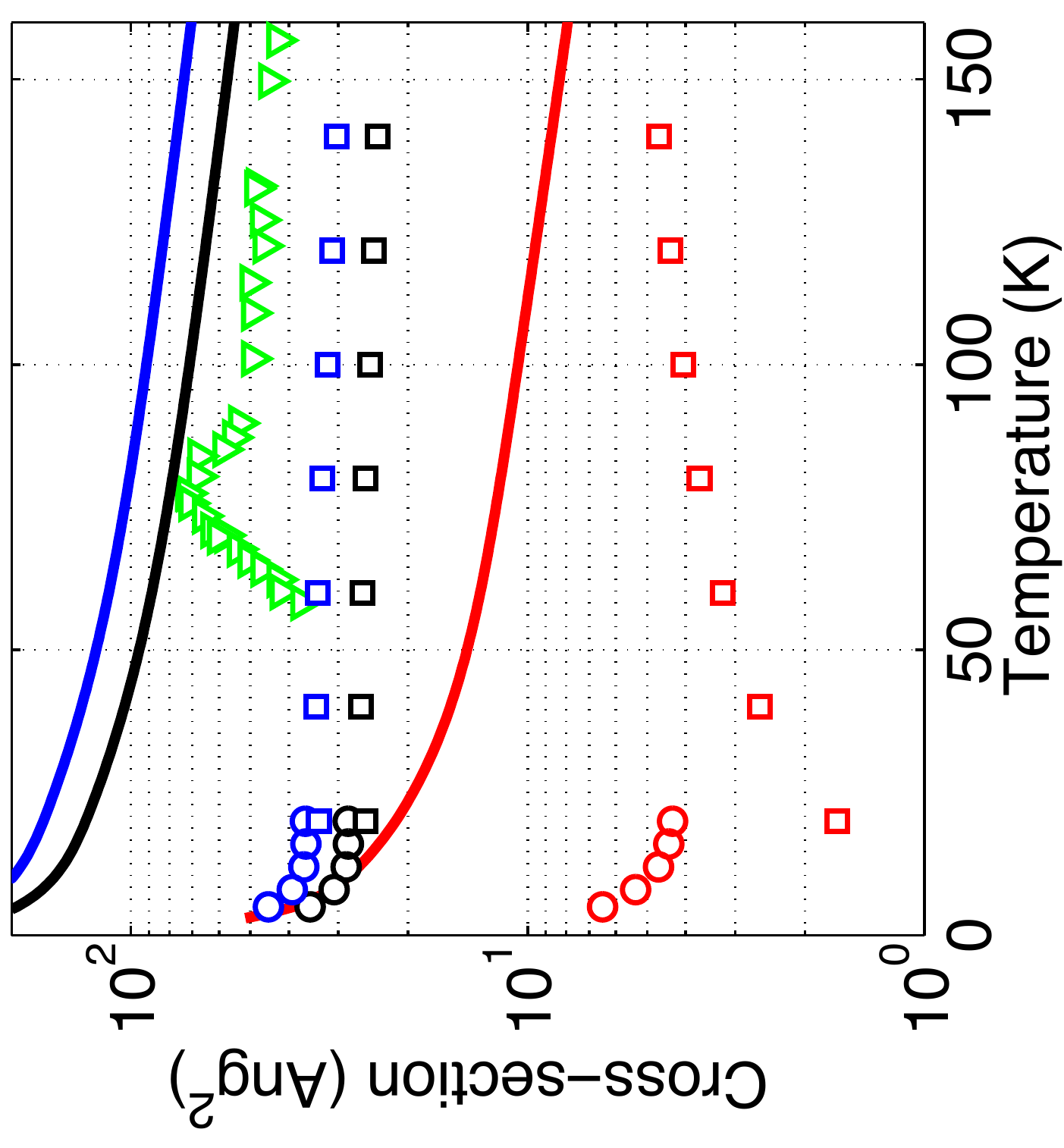}
\caption{\label{fig:comparison}Pressure broadening cross-sections for
  the transition at 556~GHz, ortho-H$_2$O, $1_{10}\leftarrow
  1_{01}$. Blue, red and black continuous lines, present theories, as
  in figure \protect\ref{fig:985}. Open symbols, summation of
  inelastic scattering cross sections as in DDP10, circles from
  Dubernet et al., 2002, squares, from Green et al., 1996, with color
  scheme identical to the solid lines. Green triangles, experimental
  values taken from DDP10.}
\end{center}
\end{figure}

We show in figure \ref{fig:comparison} the present results as compared
to estimates based on the inelastic sum, Eq.~(4), but truncated to the
first two terms, as used by DDP10 (and corrected for an error of a
factor of 2 for the Dubernet et al. data). We see that there is a
strong disagreement between the two computational schemes. In
particular, data computed from the inelastic rates of Dubernet et
al. \cite{2006A&A...460..323D}, which are based on the same PES as the
present work, are significantly lower than the present rigorous
calculations. The observed differences are to be attributed to the
approximation in applying the Random Phase Approximation (neglect of
the elastic scattering interference term) to
Eq.~(\ref{eq:optical}). Furthermore, DDP10 made the further
simplification of replacing the averaging procedure of
Eq.~\ref{eq:ave} by using rate coefficients divided by the average
thermal velocity. While this should not change the trend of the
$\spb(T)$, it adds some further imprecision. For $T>20$~K, the DDP10
points were extracted, via the same procedure, from the older values
of Phillips et al. \cite{1996ApJS..107..467P} based on a less accurate
PES. We note that the same procedure was also applied by DDP10 to the
case of H$_2$O$-$He where it was found to be quite accurate,
suggesting a very different structure of the scattering
amplitudes. This is actually not surprising since {\it i)} the
H$_2$O-He and H$_2$O-H$_2$ PES are very different (see, e.g.,
\cite{2010a:yang,2010b:yang}) and {\it ii)} additional coupling terms
are introduced by the rotational angular momentum of H$_2$.

The low temperature range of the theory vs. experiment comparison, $T
\lesssim 80$~K, is more problematic. Experiments show a dramatic
\emph{decrease} of $ \spb(T)$ as $T$ goes below about 80~K, for both
transitions examined here and also for higher transitions. An
analogous, even if less pronounced effect was also found for $\spb(T)$
in \cite{2005JChPh.122r4319R,2007JMoSp.246..118T}, for HCN-He and
$^{13}$CO-He respectively, albeit at much lower $T$, around 5K. No
definite explanation may be found in these preliminary calculations,
but it must be noted that disagreement between computations and
experiments arise at energies where the pressure broadening and
inelastic cross sections enter in a regime where narrow resonances
become prominent, see e.g.\cite{2002A&A...390..793D}. If resonances are
sufficiently narrow, that is, if the complex $\mathrm{H_2O - H_2}$ is
sufficiently long-lived, the impact approximation may be no longer
valid. According to the density in the DDP010 experiment ($n \sim
10^{20} \,\mathrm{cm}^{-3}$, B. Drouin, private communication), and
with a cross-section of about $\spb(E=80\,\mathrm{K})\simeq 50
$\AA$^2$, an average speed of $v=\sqrt{2E/\mu}\simeq 1 $~km/sec, and a
resonance width of about $\Gamma \simeq 1\,\mathrm{cm}^{-1}$
\cite{2006A&A...460..323D}, we have that the interaction volume $U$
becomes comparable to the inverse density \cite{1958PhRv..111..481B}~:
\begin{equation}
 U = (hv\spb/\Gamma)\simeq 1.6\,10^{-20}\,\mathrm{cm}^{3} \sim 1/n \simeq  10^{-20}\,\mathrm{cm}^{3}\label{eq:times},
\end{equation}
invalidating the impact approximation. It is also possible that, for
yet unknown reasons, there is a dramatic conversion from ortho-\hh\ to
para-\hh\ at low temperatures, through some paramagnetic impurities in
the walls of the cell \cite{helium:1988}, even if there is no
experimental hint that indicates this explanation at the moment. Further insight
is obviously needed in those regimes. Measurements with
para-H$_2$($j=0$) would be particularly valuable both for comparison
with theory and for applications to cold interstellar clouds where
H$_2$ is mostly in its para form \cite{2009A&A...506.1243T}.

\par In this Communication, we have shown using full quantum
scattering methods combined with state-of-art PES, that  very good
agreement is found between experimental pressure broadening and
theoretical calculations at temperatures where the impact
approximation is clearly valid, i.e. $T\gtrsim 80$~K. We have found that
the rigorous theory of Baranger is needed to make valuable
predictions for the present system and that simple approximations
based on the manipulation of inelastic rates or cross sections are
unreliable. We stress in particular that the sharp drop observed in
the pressure broadening measurements below $\sim 80$~K, and not
predicted by the present calculations, does not cast doubt on the
inelastic rates computed by Dubernet et
al. \cite{2006A&A...460..323D,2009A&A...497..911D,2010A&A...517A..13D} since a rigorous
quantum theory of broadening beyond the impact approximation seems
necessary in this regime, as discussed above. As a result, to our
opinion, the water-hydrogen V08 PES is once more successfully tested,
with a special emphasis on the mid- to long-range region of the PES
and the isotropic part. We thus complement here various tests
performed recently on the water-hydrogen system, such as differential
measurements \cite{2010b:yang}, and molecular beam scattering
experiments \cite{belpassi10}, which have all so far
confirmed the high accuracy of the V08 PES. Another series of
experiments, now underway, would aim at the spectroscopy of the bound
\hho-\hh\ van der Waals molecule. With all these tests completed in the
near future, the V08 PES will be thoroughly tested and extremely good
confidence should be gained for all astrophysical applications.
\begin{acknowledgments}
We thank B. Drouin for communicating detailed experimental data as
well as F. Thibault for useful insights. This project is partly
supported by the Institut National des Sciences de l'Univers through
its program Physico-Chimie de la Mati\`ere Interstellaire and by the
CNES. The Laboratoire d'Astrophysique is a joint CNRS/Universit\'e
Joseph-Fourier institute, under the name UMR~5585.
\end{acknowledgments}

\begin{thebibliography}{36}
\expandafter\ifx\csname natexlab\endcsname\relax\def\natexlab#1{#1}\fi
\expandafter\ifx\csname bibnamefont\endcsname\relax
  \def\bibnamefont#1{#1}\fi
\expandafter\ifx\csname bibfnamefont\endcsname\relax
  \def\bibfnamefont#1{#1}\fi
\expandafter\ifx\csname citenamefont\endcsname\relax
  \def\citenamefont#1{#1}\fi
\expandafter\ifx\csname url\endcsname\relax
  \def\url#1{\texttt{#1}}\fi
\expandafter\ifx\csname urlprefix\endcsname\relax\def\urlprefix{URL }\fi
\providecommand{\bibinfo}[2]{#2}
\providecommand{\eprint}[2][]{\url{#2}}

\bibitem[{\citenamefont{{Cernicharo} and
  {Crovisier}}(2005)}]{2005SSRv..119...29C}
\bibinfo{author}{\bibfnamefont{J.}~\bibnamefont{{Cernicharo}}}
  \bibnamefont{and}
  \bibinfo{author}{\bibfnamefont{J.}~\bibnamefont{{Crovisier}}},
  \bibinfo{journal}{Space Science Reviews} \textbf{\bibinfo{volume}{119}},
  \bibinfo{pages}{29} (\bibinfo{year}{2005}).

\bibitem[{\citenamefont{Elitzur}(1992)}]{eltzur:book}
\bibinfo{author}{\bibfnamefont{M.}~\bibnamefont{Elitzur}},
  \emph{\bibinfo{title}{Astronomical Masers}} (\bibinfo{publisher}{Springer},
  \bibinfo{address}{New York}, \bibinfo{year}{1992}).

\bibitem[{\citenamefont{Yang et~al.}(2010{\natexlab{a}})\citenamefont{Yang,
  Sarma, ter Meulen, Parker, Buck, and Wiesenfeld}}]{2010a:yang}
\bibinfo{author}{\bibfnamefont{C.-H.} \bibnamefont{Yang}},
  \bibinfo{author}{\bibfnamefont{G.}~\bibnamefont{Sarma}},
  \bibinfo{author}{\bibfnamefont{J.}~\bibnamefont{ter Meulen}},
  \bibinfo{author}{\bibfnamefont{D.~H.} \bibnamefont{Parker}},
  \bibinfo{author}{\bibfnamefont{U.}~\bibnamefont{Buck}}, \bibnamefont{and}
  \bibinfo{author}{\bibfnamefont{L.}~\bibnamefont{Wiesenfeld}},
  \bibinfo{journal}{Journal of Physical Chemistry A}
  \textbf{\bibinfo{volume}{In press}} (\bibinfo{year}{2010}{\natexlab{a}}).

\bibitem[{\citenamefont{Yang et~al.}(2010{\natexlab{b}})\citenamefont{Yang,
  Sarma, ter Meulen, Parker, McBane, Wiesenfeld, Faure, Scribano, and
  Feautrier}}]{2010b:yang}
\bibinfo{author}{\bibfnamefont{C.-H.} \bibnamefont{Yang}},
  \bibinfo{author}{\bibfnamefont{G.}~\bibnamefont{Sarma}},
  \bibinfo{author}{\bibfnamefont{J.}~\bibnamefont{ter Meulen}},
  \bibinfo{author}{\bibfnamefont{D.~H.} \bibnamefont{Parker}},
  \bibinfo{author}{\bibfnamefont{G.~C.} \bibnamefont{McBane}},
  \bibinfo{author}{\bibfnamefont{L.}~\bibnamefont{Wiesenfeld}},
  \bibinfo{author}{\bibfnamefont{A.}~\bibnamefont{Faure}},
  \bibinfo{author}{\bibfnamefont{Y.}~\bibnamefont{Scribano}}, \bibnamefont{and}
  \bibinfo{author}{\bibfnamefont{N.}~\bibnamefont{Feautrier}},
  \bibinfo{journal}{Journal of Chemical Physics} \textbf{\bibinfo{volume}{In
  press}} (\bibinfo{year}{2010}{\natexlab{b}}).

\bibitem[{\citenamefont{{{Belpassi}, L. and {Reca}, M. and {Tarantelli}, F. and
  {Roncaratti}, L. and {Pirani}, F. and {Cappelletti}, D. and {Faure}, A. and
  {Scribano}, Y.}}(2010)}]{belpassi10}
\bibinfo{author}{\bibnamefont{{{Belpassi}, L. and {Reca}, M. and {Tarantelli},
  F. and {Roncaratti}, L. and {Pirani}, F. and {Cappelletti}, D. and {Faure},
  A. and {Scribano}, Y.}}}, \emph{\bibinfo{title}{{Charge-transfer energy in
  the water-hydrogen molecular aggregate revealed by molecular-beam scattering
  experiments, charge displacement analysis, and ab-initio calculations}}},
  \bibinfo{howpublished}{{J. Am. Chem. Soc.}} (\bibinfo{year}{2010}).

\bibitem[{\citenamefont{{Daniel} et~al.}(2010)\citenamefont{{Daniel},
  {Dubernet}, {Pacaud}, and {Grosjean}}}]{2010A&A...517A..13D}
\bibinfo{author}{\bibfnamefont{F.}~\bibnamefont{{Daniel}}},
  \bibinfo{author}{\bibfnamefont{M.}~\bibnamefont{{Dubernet}}},
  \bibinfo{author}{\bibfnamefont{F.}~\bibnamefont{{Pacaud}}}, \bibnamefont{and}
  \bibinfo{author}{\bibfnamefont{A.}~\bibnamefont{{Grosjean}}},
  \bibinfo{journal}{\aap} \textbf{\bibinfo{volume}{517}}, \bibinfo{pages}{A13+}
  (\bibinfo{year}{2010}).

\bibitem[{\citenamefont{{Faure} et~al.}(2005)\citenamefont{{Faure}, {Valiron},
  {Wernli}, {Wiesenfeld}, {Rist}, {Noga}, and
  {Tennyson}}}]{2005JChPh.122v1102F}
\bibinfo{author}{\bibfnamefont{A.}~\bibnamefont{{Faure}}},
  \bibinfo{author}{\bibfnamefont{P.}~\bibnamefont{{Valiron}}},
  \bibinfo{author}{\bibfnamefont{M.}~\bibnamefont{{Wernli}}},
  \bibinfo{author}{\bibfnamefont{L.}~\bibnamefont{{Wiesenfeld}}},
  \bibinfo{author}{\bibfnamefont{C.}~\bibnamefont{{Rist}}},
  \bibinfo{author}{\bibfnamefont{J.}~\bibnamefont{{Noga}}}, \bibnamefont{and}
  \bibinfo{author}{\bibfnamefont{J.}~\bibnamefont{{Tennyson}}},
  \bibinfo{journal}{\jcp} \textbf{\bibinfo{volume}{122}},
  \bibinfo{pages}{221102} (\bibinfo{year}{2005}).

\bibitem[{\citenamefont{{Valiron} et~al.}(2008)\citenamefont{{Valiron},
  {Wernli}, {Faure}, {Wiesenfeld}, {Rist}, {Ked{\v z}uch}, and
  {Noga}}}]{2008JChPh.129m4306V}
\bibinfo{author}{\bibfnamefont{P.}~\bibnamefont{{Valiron}}},
  \bibinfo{author}{\bibfnamefont{M.}~\bibnamefont{{Wernli}}},
  \bibinfo{author}{\bibfnamefont{A.}~\bibnamefont{{Faure}}},
  \bibinfo{author}{\bibfnamefont{L.}~\bibnamefont{{Wiesenfeld}}},
  \bibinfo{author}{\bibfnamefont{C.}~\bibnamefont{{Rist}}},
  \bibinfo{author}{\bibfnamefont{S.}~\bibnamefont{{Ked{\v z}uch}}},
  \bibnamefont{and} \bibinfo{author}{\bibfnamefont{J.}~\bibnamefont{{Noga}}},
  \bibinfo{journal}{\jcp} \textbf{\bibinfo{volume}{129}},
  \bibinfo{pages}{134306} (\bibinfo{year}{2008}).

\bibitem[{\citenamefont{{Dick} et~al.}(2009)\citenamefont{{Dick}, {Drouin}, and
  {Pearson}}}]{2009JQSRT.110..619D}
\bibinfo{author}{\bibfnamefont{M.~J.} \bibnamefont{{Dick}}},
  \bibinfo{author}{\bibfnamefont{B.~J.} \bibnamefont{{Drouin}}},
  \bibnamefont{and} \bibinfo{author}{\bibfnamefont{J.~C.}
  \bibnamefont{{Pearson}}}, \bibinfo{journal}{Journal of Quantitative
  Spectroscopy and Radiative Transfer} \textbf{\bibinfo{volume}{110}},
  \bibinfo{pages}{619} (\bibinfo{year}{2009}).

\bibitem[{\citenamefont{{Dick} et~al.}(2010)\citenamefont{{Dick}, {Drouin}, and
  {Pearson}}}]{2010PhRvA..81b2706D}
\bibinfo{author}{\bibfnamefont{M.~J.} \bibnamefont{{Dick}}},
  \bibinfo{author}{\bibfnamefont{B.~J.} \bibnamefont{{Drouin}}},
  \bibnamefont{and} \bibinfo{author}{\bibfnamefont{J.~C.}
  \bibnamefont{{Pearson}}}, \bibinfo{journal}{\pra}
  \textbf{\bibinfo{volume}{81}}, \bibinfo{pages}{022706}
  (\bibinfo{year}{2010}).

\bibitem[{\citenamefont{{Thibault} et~al.}(2008)\citenamefont{{Thibault},
  {Corretja}, {Viel}, {Bermejo}, {Mart{\'{\i}}nez}, and
  {Bussery-Honvault}}}]{2008PCCP...10.5419T}
\bibinfo{author}{\bibfnamefont{F.}~\bibnamefont{{Thibault}}},
  \bibinfo{author}{\bibfnamefont{B.}~\bibnamefont{{Corretja}}},
  \bibinfo{author}{\bibfnamefont{A.}~\bibnamefont{{Viel}}},
  \bibinfo{author}{\bibfnamefont{D.}~\bibnamefont{{Bermejo}}},
  \bibinfo{author}{\bibfnamefont{R.~Z.} \bibnamefont{{Mart{\'{\i}}nez}}},
  \bibnamefont{and}
  \bibinfo{author}{\bibfnamefont{B.}~\bibnamefont{{Bussery-Honvault}}},
  \bibinfo{journal}{Physical Chemistry Chemical Physics (Incorporating Faraday
  Transactions)} \textbf{\bibinfo{volume}{10}}, \bibinfo{pages}{5419}
  (\bibinfo{year}{2008}).

\bibitem[{\citenamefont{{Thibault} et~al.}(2007)\citenamefont{{Thibault},
  {Mantz}, {Claveau}, {Henry}, {Valentin}, and
  {Hurtmans}}}]{2007JMoSp.246..118T}
\bibinfo{author}{\bibfnamefont{F.}~\bibnamefont{{Thibault}}},
  \bibinfo{author}{\bibfnamefont{A.~W.} \bibnamefont{{Mantz}}},
  \bibinfo{author}{\bibfnamefont{C.}~\bibnamefont{{Claveau}}},
  \bibinfo{author}{\bibfnamefont{A.}~\bibnamefont{{Henry}}},
  \bibinfo{author}{\bibfnamefont{A.}~\bibnamefont{{Valentin}}},
  \bibnamefont{and}
  \bibinfo{author}{\bibfnamefont{D.}~\bibnamefont{{Hurtmans}}},
  \bibinfo{journal}{Journal of Molecular Spectroscopy}
  \textbf{\bibinfo{volume}{246}}, \bibinfo{pages}{118} (\bibinfo{year}{2007}).

\bibitem[{\citenamefont{{Ronningen} and {De
  Lucia}}(2005)}]{2005JChPh.122r4319R}
\bibinfo{author}{\bibfnamefont{T.~J.} \bibnamefont{{Ronningen}}}
  \bibnamefont{and} \bibinfo{author}{\bibfnamefont{F.~C.} \bibnamefont{{De
  Lucia}}}, \bibinfo{journal}{\jcp} \textbf{\bibinfo{volume}{122}},
  \bibinfo{pages}{184319} (\bibinfo{year}{2005}).

\bibitem[{\citenamefont{{Green}}(1991)}]{1991JChPh..95.3888G}
\bibinfo{author}{\bibfnamefont{S.}~\bibnamefont{{Green}}},
  \bibinfo{journal}{\jcp} \textbf{\bibinfo{volume}{95}}, \bibinfo{pages}{3888}
  (\bibinfo{year}{1991}).

\bibitem[{\citenamefont{{Green} et~al.}(1991)\citenamefont{{Green}, {Defrees},
  and {McLean}}}]{1991JChPh..94.1346G}
\bibinfo{author}{\bibfnamefont{S.}~\bibnamefont{{Green}}},
  \bibinfo{author}{\bibfnamefont{D.~J.} \bibnamefont{{Defrees}}},
  \bibnamefont{and} \bibinfo{author}{\bibfnamefont{A.~D.}
  \bibnamefont{{McLean}}}, \bibinfo{journal}{\jcp}
  \textbf{\bibinfo{volume}{94}}, \bibinfo{pages}{1346} (\bibinfo{year}{1991}).

\bibitem[{\citenamefont{{Dubernet} et~al.}(2006)\citenamefont{{Dubernet},
  {Daniel}, {Grosjean}, {Faure}, {Valiron}, {Wernli}, {Wiesenfeld}, {Rist},
  {Noga}, and {Tennyson}}}]{2006A&A...460..323D}
\bibinfo{author}{\bibfnamefont{M.}~\bibnamefont{{Dubernet}}},
  \bibinfo{author}{\bibfnamefont{F.}~\bibnamefont{{Daniel}}},
  \bibinfo{author}{\bibfnamefont{A.}~\bibnamefont{{Grosjean}}},
  \bibinfo{author}{\bibfnamefont{A.}~\bibnamefont{{Faure}}},
  \bibinfo{author}{\bibfnamefont{P.}~\bibnamefont{{Valiron}}},
  \bibinfo{author}{\bibfnamefont{M.}~\bibnamefont{{Wernli}}},
  \bibinfo{author}{\bibfnamefont{L.}~\bibnamefont{{Wiesenfeld}}},
  \bibinfo{author}{\bibfnamefont{C.}~\bibnamefont{{Rist}}},
  \bibinfo{author}{\bibfnamefont{J.}~\bibnamefont{{Noga}}}, \bibnamefont{and}
  \bibinfo{author}{\bibfnamefont{J.}~\bibnamefont{{Tennyson}}},
  \bibinfo{journal}{\aap} \textbf{\bibinfo{volume}{460}}, \bibinfo{pages}{323}
  (\bibinfo{year}{2006}).

\bibitem[{\citenamefont{{Baranger}}(1958{\natexlab{a}})}]{1958PhRv..112..855B}
\bibinfo{author}{\bibfnamefont{M.}~\bibnamefont{{Baranger}}},
  \bibinfo{journal}{Physical Review} \textbf{\bibinfo{volume}{112}},
  \bibinfo{pages}{855} (\bibinfo{year}{1958}{\natexlab{a}}).

\bibitem[{\citenamefont{{Ben-Reuven}}(1975)}]{1975AdChP..33..235B}
\bibinfo{author}{\bibfnamefont{A.}~\bibnamefont{{Ben-Reuven}}},
  \bibinfo{journal}{Advances in Chemical Physics}
  \textbf{\bibinfo{volume}{33}}, \bibinfo{pages}{235} (\bibinfo{year}{1975}).

\bibitem[{\citenamefont{{Baranger}}(1958{\natexlab{b}})}]{1958PhRv..111..481B}
\bibinfo{author}{\bibfnamefont{M.}~\bibnamefont{{Baranger}}},
  \bibinfo{journal}{Physical Review} \textbf{\bibinfo{volume}{111}},
  \bibinfo{pages}{481} (\bibinfo{year}{1958}{\natexlab{b}}).

\bibitem[{\citenamefont{{Schaefer} and {Monchick}}(1992)}]{1992A&A...265..859S}
\bibinfo{author}{\bibfnamefont{J.}~\bibnamefont{{Schaefer}}} \bibnamefont{and}
  \bibinfo{author}{\bibfnamefont{L.}~\bibnamefont{{Monchick}}},
  \bibinfo{journal}{\aap} \textbf{\bibinfo{volume}{265}}, \bibinfo{pages}{859}
  (\bibinfo{year}{1992}).

\bibitem[{\citenamefont{{Schaefer} and {Monchick}}(1987)}]{1987JChPh..87..171S}
\bibinfo{author}{\bibfnamefont{J.}~\bibnamefont{{Schaefer}}} \bibnamefont{and}
  \bibinfo{author}{\bibfnamefont{L.}~\bibnamefont{{Monchick}}},
  \bibinfo{journal}{\jcp} \textbf{\bibinfo{volume}{87}}, \bibinfo{pages}{171}
  (\bibinfo{year}{1987}).

\bibitem[{\citenamefont{{Green}}(1977)}]{1977CPL:green}
\bibinfo{author}{\bibfnamefont{S.}~\bibnamefont{{Green}}},
  \bibinfo{journal}{Chemical Physics Letters} \textbf{\bibinfo{volume}{47}},
  \bibinfo{pages}{119} (\bibinfo{year}{1977}).

\bibitem[{\citenamefont{{Thibault} et~al.}(2001)\citenamefont{{Thibault},
  {Calil}, {Buldyreva}, {Chrysos}, {Hartmann}, and
  {Bouanich}}}]{2001PCCP....3.3924T}
\bibinfo{author}{\bibfnamefont{F.}~\bibnamefont{{Thibault}}},
  \bibinfo{author}{\bibfnamefont{B.}~\bibnamefont{{Calil}}},
  \bibinfo{author}{\bibfnamefont{J.}~\bibnamefont{{Buldyreva}}},
  \bibinfo{author}{\bibfnamefont{M.}~\bibnamefont{{Chrysos}}},
  \bibinfo{author}{\bibfnamefont{J.}~\bibnamefont{{Hartmann}}},
  \bibnamefont{and}
  \bibinfo{author}{\bibfnamefont{J.}~\bibnamefont{{Bouanich}}},
  \bibinfo{journal}{Physical Chemistry Chemical Physics (Incorporating Faraday
  Transactions)} \textbf{\bibinfo{volume}{3}}, \bibinfo{pages}{3924}
  (\bibinfo{year}{2001}).

\bibitem[{\citenamefont{Green et~al.}(1977)\citenamefont{Green, Monchick,
  Goldflam, and Kouri}}]{1977JCPgreen:1409}
\bibinfo{author}{\bibfnamefont{S.}~\bibnamefont{Green}},
  \bibinfo{author}{\bibfnamefont{L.}~\bibnamefont{Monchick}},
  \bibinfo{author}{\bibfnamefont{R.}~\bibnamefont{Goldflam}}, \bibnamefont{and}
  \bibinfo{author}{\bibfnamefont{D.~J.} \bibnamefont{Kouri}},
  \bibinfo{journal}{The Journal of Chemical Physics}
  \textbf{\bibinfo{volume}{66}}, \bibinfo{pages}{1409} (\bibinfo{year}{1977}),
  \urlprefix\url{http://link.aip.org/link/?JCP/66/1409/1}.

\bibitem[{\citenamefont{{Hutson} and {Green}}(1994)}]{molscat}
\bibinfo{author}{\bibfnamefont{J.~M.} \bibnamefont{{Hutson}}} \bibnamefont{and}
  \bibinfo{author}{\bibfnamefont{S.}~\bibnamefont{{Green}}},
  \emph{\bibinfo{title}{MOLSCAT computer code, version 14, distributed by
  Collaborative Computational Project No. 6 of the Engineering and Physical
  Sciences Research Council UK),}} (\bibinfo{year}{1994}).

\bibitem[{\citenamefont{{Dubernet} et~al.}(2009)\citenamefont{{Dubernet},
  {Daniel}, {Grosjean}, and {Lin}}}]{2009A&A...497..911D}
\bibinfo{author}{\bibfnamefont{M.}~\bibnamefont{{Dubernet}}},
  \bibinfo{author}{\bibfnamefont{F.}~\bibnamefont{{Daniel}}},
  \bibinfo{author}{\bibfnamefont{A.}~\bibnamefont{{Grosjean}}},
  \bibnamefont{and} \bibinfo{author}{\bibfnamefont{C.~Y.} \bibnamefont{{Lin}}},
  \bibinfo{journal}{\aap} \textbf{\bibinfo{volume}{497}}, \bibinfo{pages}{911}
  (\bibinfo{year}{2009}).

\bibitem[{\citenamefont{{Grosjean} et~al.}(2003)\citenamefont{{Grosjean},
  {Dubernet}, and {Ceccarelli}}}]{2003A&A...408.1197G}
\bibinfo{author}{\bibfnamefont{A.}~\bibnamefont{{Grosjean}}},
  \bibinfo{author}{\bibfnamefont{M.}~\bibnamefont{{Dubernet}}},
  \bibnamefont{and}
  \bibinfo{author}{\bibfnamefont{C.}~\bibnamefont{{Ceccarelli}}},
  \bibinfo{journal}{\aap} \textbf{\bibinfo{volume}{408}}, \bibinfo{pages}{1197}
  (\bibinfo{year}{2003}).

\bibitem[{\citenamefont{{Dubernet} and {Grosjean}}(2002)}]{2002A&A...390..793D}
\bibinfo{author}{\bibfnamefont{M.}~\bibnamefont{{Dubernet}}} \bibnamefont{and}
  \bibinfo{author}{\bibfnamefont{A.}~\bibnamefont{{Grosjean}}},
  \bibinfo{journal}{\aap} \textbf{\bibinfo{volume}{390}}, \bibinfo{pages}{793}
  (\bibinfo{year}{2002}).

\bibitem[{\citenamefont{{Thibault} et~al.}(2009)\citenamefont{{Thibault},
  {Fuller}, {Grabow}, {Hardwick}, {Marcus}, {Marston}, {Robertson}, {Senning},
  {Stoffel}, and {Wiser}}}]{2009JMoSp.256...17T}
\bibinfo{author}{\bibfnamefont{F.}~\bibnamefont{{Thibault}}},
  \bibinfo{author}{\bibfnamefont{E.~P.} \bibnamefont{{Fuller}}},
  \bibinfo{author}{\bibfnamefont{K.~A.} \bibnamefont{{Grabow}}},
  \bibinfo{author}{\bibfnamefont{J.~L.} \bibnamefont{{Hardwick}}},
  \bibinfo{author}{\bibfnamefont{C.~I.} \bibnamefont{{Marcus}}},
  \bibinfo{author}{\bibfnamefont{D.}~\bibnamefont{{Marston}}},
  \bibinfo{author}{\bibfnamefont{L.~A.} \bibnamefont{{Robertson}}},
  \bibinfo{author}{\bibfnamefont{E.~N.} \bibnamefont{{Senning}}},
  \bibinfo{author}{\bibfnamefont{M.~C.} \bibnamefont{{Stoffel}}},
  \bibnamefont{and} \bibinfo{author}{\bibfnamefont{R.~S.}
  \bibnamefont{{Wiser}}}, \bibinfo{journal}{Journal of Molecular Spectroscopy}
  \textbf{\bibinfo{volume}{256}}, \bibinfo{pages}{17} (\bibinfo{year}{2009}).

\bibitem[{\citenamefont{{G{\'o}mez} et~al.}(2007)\citenamefont{{G{\'o}mez},
  {Mart{\'{\i}}nez}, {Bermejo}, {Thibault}, {Joubert}, {Bussery-Honvault}, and
  {Bonamy}}}]{2007JChPh.126t4302G}
\bibinfo{author}{\bibfnamefont{L.}~\bibnamefont{{G{\'o}mez}}},
  \bibinfo{author}{\bibfnamefont{R.~Z.} \bibnamefont{{Mart{\'{\i}}nez}}},
  \bibinfo{author}{\bibfnamefont{D.}~\bibnamefont{{Bermejo}}},
  \bibinfo{author}{\bibfnamefont{F.}~\bibnamefont{{Thibault}}},
  \bibinfo{author}{\bibfnamefont{P.}~\bibnamefont{{Joubert}}},
  \bibinfo{author}{\bibfnamefont{B.}~\bibnamefont{{Bussery-Honvault}}},
  \bibnamefont{and} \bibinfo{author}{\bibfnamefont{J.}~\bibnamefont{{Bonamy}}},
  \bibinfo{journal}{\jcp} \textbf{\bibinfo{volume}{126}},
  \bibinfo{pages}{204302} (\bibinfo{year}{2007}).

\bibitem[{\citenamefont{Hartmann et~al.}(2008)\citenamefont{Hartmann, Boulet,
  and Robert}}]{hartmann:book}
\bibinfo{author}{\bibfnamefont{J.}~\bibnamefont{Hartmann}},
  \bibinfo{author}{\bibfnamefont{C.}~\bibnamefont{Boulet}}, \bibnamefont{and}
  \bibinfo{author}{\bibfnamefont{D.}~\bibnamefont{Robert}},
  \emph{\bibinfo{title}{Collisional Effects on Molecular Spectra}}
  (\bibinfo{publisher}{Elsevier}, \bibinfo{year}{2008}).

\bibitem[{\citenamefont{{Yang} et~al.}(2008)\citenamefont{{Yang}, {Buldyreva},
  {Gordon}, {Rohart}, {Cuisset}, {Mouret}, {Bocquet}, and
  {Hindle}}}]{2008JQSRT.109.2857Y}
\bibinfo{author}{\bibfnamefont{C.}~\bibnamefont{{Yang}}},
  \bibinfo{author}{\bibfnamefont{J.}~\bibnamefont{{Buldyreva}}},
  \bibinfo{author}{\bibfnamefont{I.~E.} \bibnamefont{{Gordon}}},
  \bibinfo{author}{\bibfnamefont{F.}~\bibnamefont{{Rohart}}},
  \bibinfo{author}{\bibfnamefont{A.}~\bibnamefont{{Cuisset}}},
  \bibinfo{author}{\bibfnamefont{G.}~\bibnamefont{{Mouret}}},
  \bibinfo{author}{\bibfnamefont{R.}~\bibnamefont{{Bocquet}}},
  \bibnamefont{and} \bibinfo{author}{\bibfnamefont{F.}~\bibnamefont{{Hindle}}},
  \bibinfo{journal}{Journal of Quantitative Spectroscopy and Radiative
  Transfer} \textbf{\bibinfo{volume}{109}}, \bibinfo{pages}{2857}
  (\bibinfo{year}{2008}).

\bibitem[{\citenamefont{{Nguyen} et~al.}(2006)\citenamefont{{Nguyen},
  {Buldyreva}, {Colmont}, {Rohart}, {Wlodarczak}, and
  {Alekseev}}}]{2006MolPh.104.2701N}
\bibinfo{author}{\bibfnamefont{L.}~\bibnamefont{{Nguyen}}},
  \bibinfo{author}{\bibfnamefont{J.}~\bibnamefont{{Buldyreva}}},
  \bibinfo{author}{\bibfnamefont{J.}~\bibnamefont{{Colmont}}},
  \bibinfo{author}{\bibfnamefont{F.}~\bibnamefont{{Rohart}}},
  \bibinfo{author}{\bibfnamefont{G.}~\bibnamefont{{Wlodarczak}}},
  \bibnamefont{and}
  \bibinfo{author}{\bibfnamefont{E.}~\bibnamefont{{Alekseev}}},
  \bibinfo{journal}{Molecular Physics} \textbf{\bibinfo{volume}{104}},
  \bibinfo{pages}{2701} (\bibinfo{year}{2006}).

\bibitem[{\citenamefont{{Phillips} et~al.}(1996)\citenamefont{{Phillips},
  {Maluendes}, and {Green}}}]{1996ApJS..107..467P}
\bibinfo{author}{\bibfnamefont{T.~R.} \bibnamefont{{Phillips}}},
  \bibinfo{author}{\bibfnamefont{S.}~\bibnamefont{{Maluendes}}},
  \bibnamefont{and} \bibinfo{author}{\bibfnamefont{S.}~\bibnamefont{{Green}}},
  \bibinfo{journal}{\apjs} \textbf{\bibinfo{volume}{107}}, \bibinfo{pages}{467}
  (\bibinfo{year}{1996}).

\bibitem[{\citenamefont{{Tastevin, G.} et~al.}(1988)\citenamefont{{Tastevin,
  G.}, {Nacher, P.J.}, {Wiesenfeld, L.}, {Leduc, M.}, and {Lalo\"e,
  F.}}}]{helium:1988}
\bibinfo{author}{\bibnamefont{{Tastevin, G.}}},
  \bibinfo{author}{\bibnamefont{{Nacher, P.J.}}},
  \bibinfo{author}{\bibnamefont{{Wiesenfeld, L.}}},
  \bibinfo{author}{\bibnamefont{{Leduc, M.}}}, \bibnamefont{and}
  \bibinfo{author}{\bibnamefont{{Lalo\"e, F.}}}, \bibinfo{journal}{J. Phys.
  France} \textbf{\bibinfo{volume}{49}}, \bibinfo{pages}{1}
  (\bibinfo{year}{1988}).

\bibitem[{\citenamefont{{Troscompt} et~al.}(2009)\citenamefont{{Troscompt},
  {Faure}, {Maret}, {Ceccarelli}, {Hily-Blant}, and
  {Wiesenfeld}}}]{2009A&A...506.1243T}
\bibinfo{author}{\bibfnamefont{N.}~\bibnamefont{{Troscompt}}},
  \bibinfo{author}{\bibfnamefont{A.}~\bibnamefont{{Faure}}},
  \bibinfo{author}{\bibfnamefont{S.}~\bibnamefont{{Maret}}},
  \bibinfo{author}{\bibfnamefont{C.}~\bibnamefont{{Ceccarelli}}},
  \bibinfo{author}{\bibfnamefont{P.}~\bibnamefont{{Hily-Blant}}},
  \bibnamefont{and}
  \bibinfo{author}{\bibfnamefont{L.}~\bibnamefont{{Wiesenfeld}}},
  \bibinfo{journal}{\aap} \textbf{\bibinfo{volume}{506}}, \bibinfo{pages}{1243}
  (\bibinfo{year}{2009}).

\end{thebibliography}

\end{document}